\documentstyle[12pt]{article}
\def\be{\begin{equation}}
\def\ee{\end{equation}}
\def\bc{\begin{center}}
\def\ec{\end{center}}
\def\bea{\begin{eqnarray}}
\def\eea{\end{eqnarray}}

\def\ov{\overline}

\def\mpl{M_{\rm P}}

\def\simlt{\stackrel{<}{{}_\sim}}

\catcode`@=11
\def\marginnote#1{}
\newcount\hour
\newcount\minute
\newtoks\amorpm
\hour=\time\divide\hour by60
\minute=\time{\multiply\hour by60 \global\advance\minute by-\hour}
\edef\standardtime{{\ifnum\hour<12 \global\amorpm={am}%
        \else\global\amorpm={pm}\advance\hour by-12 \fi
        \ifnum\hour=0 \hour=12 \fi
        \number\hour:\ifnum\minute<10 0\fi\number\minute\the\amorpm}}
\edef\militarytime{\number\hour:\ifnum\minute<10 0\fi\number\minute}
\def\draftlabel#1{{\@bsphack\if@filesw {\let\thepage\relax
   \xdef\@gtempa{\write\@auxout{\string
      \newlabel{#1}{{\@currentlabel}{\thepage}}}}}\@gtempa
   \if@nobreak \ifvmode\nobreak\fi\fi\fi\@esphack}
        \gdef\@eqnlabel{#1}}
\def\@eqnlabel{}
\def\@vacuum{}
\def\draftmarginnote#1{\marginpar{\raggedright\scriptsize\tt#1}}
\def\draft{\oddsidemargin 0.0truein
        \def\@oddfoot{\sl preliminary draft \hfil
        \rm\thepage\hfil\sl\today\quad\militarytime}
        \let\@evenfoot\@oddfoot \overfullrule 3pt
        \let\label=\draftlabel
        \let\marginnote=\draftmarginnote
   \def\@eqnnum{(\theequation)\rlap{\kern\marginparsep\tt\@eqnlabel}%
\global\let\@eqnlabel\@vacuum}  }
\catcode`@=12
\begin{document}
\begin{titlepage}
\vspace*{-1cm}
\phantom{bla}
\hfill{CERN-TH.7439/94}
\\
\phantom{bla}
\hfill{LBL-35117}
\\
\phantom{bla}
\hfill{hep-th/9409099}
\vskip 1.5cm
\begin{center}
{\Large\bf Supersymmetry and $SU(2) \times U(1)$ breaking
\\
with naturally vanishing vacuum energy}
\footnote{Work supported in part by the US NSF under grant
No.~PHY89--04035, by the US DOE under Contract DE-AC03-76SF00098,
and by the European Union under contract No.~CHRX-CT92-0004.}
\end{center}
\vskip 1.0cm
\begin{center}
{\large Andrea Brignole}\footnote{Supported by an INFN Postdoctoral
Fellowship.} \\
\vskip .1cm
Lawrence Berkeley Laboratory, Berkeley CA 94720, USA
\\
\vskip .2cm
and
\\
\vskip .2cm
{\large Fabio
Zwirner}\footnote{On leave from INFN, Sezione di Padova, Padua,
Italy.}
\\
\vskip .1cm
Theory Division, CERN, CH-1211 Geneva 23, Switzerland
\end{center}
\vskip 0.5cm
\begin{abstract}
\noindent
We show how the spontaneous breaking of local $N=1$ supersymmetry and
of the $SU(2) \times U(1)$ gauge symmetry can be simultaneously
realized, with naturally vanishing tree-level vacuum energy, in
superstring
effective supergravities. Both the gravitino mass $m_{3/2}$ and the
electroweak scale $m_Z$ are classically undetermined, and slide along
moduli directions that include the Higgs flat direction $|H_1^0| =
|H_2^0|$.  There are important differences with conventional
supergravity models: the goldstino has components along the higgsino
direction; $SU(2) \times U(1)$ breaking occurs already at the
classical
level; the scales $m_{3/2}$ and $m_Z$, the gauge couplings, the
lightest
Higgs mass and the cosmological constant are entirely determined by
quantum corrections.
\end{abstract}
\vfill{
CERN-TH.7439/94
\newline
\noindent
September 1994}
\end{titlepage}
\setcounter{footnote}{0}
\vskip2truecm
\vspace{1cm}
{\bf 1.}
The gauge hierarchy problem of the Standard Model (SM), linked to the
fact that the natural mass scale for an elementary Higgs field is the
ultraviolet cutoff, points towards supersymmetric extensions of the
SM [\ref{susy}] for describing particle interactions above the
electroweak scale. Along this direction, the Minimal Supersymmetric
Standard Model (MSSM) should not be considered more than a plausible
phenomenological parametrization, possibly useful for organizing
experimental searches. A theoretically satisfactory solution of the
gauge
hierarchy problem requires a model for spontaneous supersymmetry
breaking within a more fundamental theory, such as $N=1$
supergravity, seen in turn as the low-energy limit of a consistent
quantum theory at the Planck scale. Besides the gauge hierarchy
problem, related to the smallness of the ratio $m_Z/\mpl \sim
10^{-16}$, in spontaneously broken supergravity there is a second
hierarchy problem, associated with the vacuum energy\footnote{For a
review of the cosmological constant problem, see e.g.
ref.~[\ref{weinberg}].}: since the potential is not
positive-semidefinite, the natural scale for its VEV after
supersymmetry breaking is $\langle V \rangle = {\cal O} (m_{3/2}^2
\mpl^2)$, and one has to explain why it is not so, but instead, after
the electroweak and other phase transitions, $\langle V \rangle /
\mpl^4 \simlt 10^{-120}$.

Despite intense theoretical efforts over more than a decade, it is
not
yet clear how the above problems could find a solution. If one tries
to make sense of perturbation theory around a flat classical
background, avoiding unnatural fine-tunings, a promising starting
point are the supergravity models [\ref{nscl},\ref{nsqu}]
characterized by a manifestly positive-semidefinite classical
potential, with all minima corresponding to broken supersymmetry and
vanishing vacuum energy, and the gravitino mass sliding along some
flat direction, parametrized by a gauge-singlet scalar field. At the
classical level, the gravitino mass $m_{3/2}$ is undetermined and
$SU(2) \times U(1)$ is unbroken, thus the structure of quantum
corrections becomes of crucial importance for the viability of such a
scenario. At the pure supergravity level, there is no way of
controlling ${\cal O} (m_{3/2}^2 \mpl^2)$ contributions to the
effective potential, coming from loop corrections in the underlying
quantum theory of gravity, which if present would generically forbid
both the desired hierarchies. One can at most assume that these
contributions are absent, in which case [\ref{nsqu}] logarithmic
quantum corrections can induce a gravitino mass $m_{3/2} \ll \mpl$
and break $SU(2) \times U(1)$, with $m_Z = {\cal O} (m_{3/2})$. Along
this line, some progress has recently been made in the framework of
four-dimensional superstrings, where quantum corrections can be
consistently computed and incorporated in the effective supergravity
theories: it was possible to identify a restricted class of
models [\ref{lhc}] where loop contributions to the vacuum energy are
computed to be at most ${\cal O} (m_{3/2}^4)$; however, as we shall
describe below, a number of problems were still left unsolved.

To prepare for the following discussion, we would like to recall
that, in the above as well as in many other supergravity models, one
first discusses supersymmetry breaking in a classical effective
supergravity, with a gauge-singlet goldstino belonging to a hidden
sector, coupled to the MSSM states via interactions of gravitational
strength. Then one takes the flat limit, formally decoupling the
hidden sector, and recovers the MSSM with specified forms of its mass
parameters. Finally, one computes quantum corrections due to
renormalizable MSSM interactions, to discuss the radiative breaking
of $SU(2) \times U(1)$ and the dynamical determination of mass
scales.

The most important among the unsolved problems is that of the vacuum
energy, which cannot be avoided if one claims to be discussing the
low-energy effective theory of a fundamental theory of all
interactions, including the gravitational ones. Even if, in the
framework of the MSSM, one parametrizes [\ref{kpz}] the hidden sector
contribution to the cosmological constant with a potential term
$\Delta V_{cosm} = \eta m_{3/2}^4$, this is not adequate to ensure
the vanishing of the vacuum energy generated by the breaking of
$SU(2) \times U(1)$: the gravitino mass is a dynamical variable, and
the minimization condition with respect to $m_{3/2}$, $(\partial V /
\partial m_{3/2}) = 0$, is not necessarily compatible with the
condition of vanishing vacuum energy, $V=0$. In general, one expects
a non-vanishing vacuum energy ${\cal O} (m_Z^4)$ to be generated,
which is unlikely to be cancelled by phase transitions involving much
lower mass scales.

Another theoretical, more technical problem of the previous approach
is the consistent inclusion of quantum corrections. In principle,
these quantum corrections must be computed in the fundamental
four-dimensional string theory. Unfortunately, only a very restricted
class of models for supersymmetry breaking has been consistently
formulated at the string level: for the moment, we are bound to the
orbifold models in which supersymmetry is spontaneously broken at the
string tree level [\ref{ss}]. Moreover, despite many recent
developments [\ref{strqu}] in the computation of string loop
corrections to the defining functions of the low-energy effective
supergravity, complete results, including all the relevant
field-dependences, are not yet available. The situation is even worse
for the models of supersymmetry breaking involving non-perturbative
phenomena: until now, some models based on gaugino condensation have
been formulated only at the level of the effective supergravity
theories [\ref{gcond}], whilst others [\ref{others}] have been
thoroughly discussed only at the level of global supersymmetry.

Finally, there is a set of more phenomenological cosmological
problems [\ref{cosmo}], associated with gravitationally interacting
particles with masses at the electroweak scale or below (the
gravitino and the spin~0 and the spin~$1/2$ components of some
singlet moduli fields). Such particles may overclose the universe
if they are stable, whereas they may alter the light elements
abundance if they decay after primordial nucleosynthesis. For spin~0
fields, there is the additional problem of the energy density stored
in the coherent oscillations of the classical fields, which can
exceed the closure density if it is not dissipated fast enough.

In view of the above problems, it might be interesting to look for
supergravity models in which supersymmetry and $SU(2) \times U(1)$
are both spontaneously broken at the classical level, with naturally
vanishing vacuum energy\footnote{The combined breaking of
supersymmetry
and of a grand-unified gauge symmetry was previously considered in
[\ref{quiros}].}. Once realized at the string level, they
could be the starting point for a systematic investigation of
perturbative quantum corrections, searching for symmetry properties
that might allow for the desired values of $m_Z/\mpl$ and $\langle V
\rangle / \mpl^4$. In particular, one could envisage situations in
which the goldstino has significant components along the higgsino
directions: this would produce a weakly interacting gravitino, as
originally discussed in [\ref{fayet}], with highly non-standard
cosmological properties. In these
models, a linear combination of the two MSSM Higgs doublets would act
as a modulus field, associated by supersymmetry with the goldstino;
the relation between the lightest Higgs mass and the other
supersymmetry-breaking masses might be non-trivial, due the fact that
the Higgs can be interpreted as the pseudo-Goldstone boson of some
approximate non-compact global symmetry; perhaps
this could allow for a spectrum of superpartners significantly above
the electroweak scale.

In the rest of this paper, we shall introduce a simple,
superstring-derived supergravity model that exhibits some of these
properties. We shall discuss the main features of its mass spectrum,
and show how a special version of the MSSM can be recovered in an
appropriate limit. We shall finally comment on quantum corrections
and on other loose ends of the model, suggesting some possible
improvements.

\vspace{1cm}
{\bf 2.}
We consider here the supergravity model defined by\footnote{We use
the standard supergravity mass units where $\mpl \equiv G_N^{-1/2} /
\sqrt{8 \pi} = 1$.} the gauge group $G_0 \equiv SU(3) \times
SU(2) \times U(1)$, with gauge kinetic function
\be
\label{fab}
f_{ab} = \delta_{ab} S \, ,
\ee
by the K\"ahler potential
\be
\label{kah}
K =
- \log ( S + \ov{S} )
- \log Y
+ z^{\alpha} \ov{z}_{\alpha} \, ,
\ee
where
\be
\label{yyy}
Y =
( T + \ov{T})^2
-
( H_1^0 + \ov{H_2^0} ) ( \ov{H_1^0} + H_2^0 )
+ \ldots \, ,
\ee
and by the superpotential
\be
\label{www}
w = k
+ {1 \over 2} h_{\alpha \beta}^{(1)} z^{\alpha} z^{\beta} H_1^0
+ {1 \over 2} h_{\alpha \beta}^{(2)} z^{\alpha} z^{\beta} H_2^0
+ \ldots \, ,
\ee
where for simplicity we assume the constants $k$, $h_{\alpha
\beta}^{(1)}$ and $h_{\alpha \beta}^{(2)}$ to be real.

On the one hand, this model can be seen as a locally supersymmetric
extension of the SM: the superfields $z^{\alpha}$ can be identified
with the quarks and leptons of the MSSM, the fields $(H_1^0,H_2^0)$
with the neutral components  of the supersymmetric Higgs
doublets. The $G_0$-invariant completions of the couplings in
eqs.~(\ref{yyy}) and (\ref{www}), involving the charged Higgs fields
$(H_1^-,H_2^+)$ and denoted by dots, can be trivially
worked out, but will not play an important role in the following
discussion, so we prefer to omit them for the moment.
Equation~(\ref{fab})
tells us that, at the classical level, $g_3^2 = g_2^2 = g_1^2 = (
{\rm Re}
\, S )^{-1}$, as is the case in many four-dimensional superstring
models\footnote{The conventionally normalized $U(1)$ coupling is
given, as usual, by $g'\,^2 = (3/5) g_1^2$.}. The differences among
the gauge couplings at the electroweak scale are due to quantum
corrections, and will be ignored here.

On the other hand, this model exhibits some remarkable properties of
the classical effective supergravities [\ref{efft}] corresponding to
four-dimensional superstrings [\ref{fds}]. Indeed, it can be
identified with a consistent truncation of the string model with
spontaneously broken $N=1$ supersymmetry discussed in [\ref{fkpz}]
(flat directions that break the gauge symmetry were not
explicitly investigated there, whereas here we are putting to zero
some
extra fields that play no role in the symmetry-breaking
phenomena under consideration). The singlet fields $S$ and $T$ can be
identified with some of the moduli fields: $S$ is the universal
dilaton-axion multiplet, parametrizing the K\"ahler manifold
$SU(1,1)/U(1)$; $T$ can be identified with the diagonal combination
of the $(1,1)$ and $(1,2)$ moduli associated with one complex
internal
dimension, and $T (\alpha')^{1/2}$ should not be taken too close to
$1$
for eqs.~(\ref{fab})--(\ref{www}) to be a good approximation. As
in fermionic constructions [\ref{ferm}] and in some orbifold
models [\ref{orb}], the fields
$(T,H_1^0,H_2^0,\ldots)$, transforming in real representations
of $G_0$, parametrize an $SO(2,n)/[SO(2) \times SO(n)]$ K\"ahler
manifold: the gauge-invariant parametrization for the K\"ahler
potential of the Higgs fields is taken from [\ref{param}] and can be
obtained from the one of [\ref{fkpz}], $Y = (x_0 + \ov{x}_0)^2 -
\sum_{i=1}^{n-1} (x_i + \ov{x}_i)^2$, by making the field
redefinitions
$T = x_0$, $H_1^0=x_1 + i x_2$, $H_2^0=x_1 - i x_2$, \ldots. For the
fields $z^{\alpha}$, transforming in chiral representations of $G_0$,
we
shall consider only small fluctuations around $\langle z \rangle =
0$, so that according to [\ref{fkpz}]
we can consistently assume canonical kinetic terms. As for the
superpotential, the cubic couplings are typical of a large class of
four-dimensional string models, also in the limit of exact
supersymmetry, whereas the constant $k$ is the peculiar source of
spontaneous supersymmetry breaking in the string constructions of the
type considered in [\ref{ss}]. At the level of the classical
effective
theory, it is not restrictive to choose $k=1$, since this amounts to
a trivial rescaling of the fields $T$, $H_1^0$ and $H_2^0$.

The scalar potential of the model under consideration can be computed
from the general expression
\be
V = V_F + V_D
= e^G \left[ G^i \left( G^{-1} \right)_i^{\;\; j} G_j -3 \right]
+ \frac{1}{2} \left( {\rm Re} \, f^{-1} \right)_{ab}
\left[ G^i \left( T^a \right)_i^{\;\; j} \phi_j \right]
\left[ G^k \left( T^b \right)_k^{\;\; l} \phi_l \right] \, ,
\ee
where $G \equiv K + \log|w|^2$, $\phi^i \equiv ( S, T, H_1^0, H_2^0,
\ldots, z^{\alpha})$, and we use standard supergravity conventions on
derivatives. After some simple calculation, and keeping only terms
up to second order in the fields $z^{\alpha}$, we find
\bea
V_F & = & \frac{k^2}{(S+\overline{S})Y}
\left\{ |z^{\alpha}|^2 + \left| \frac{1}{k} \left( h^{(1)}_{\alpha
\beta} H_1^0 +
h^{(2)}_{\alpha \beta} H_2^0 \right) z^{\beta} \right|^2 \right.
\nonumber \\
& & \left. +
\frac{1}{2k} \left[ \left( h^{(1)}_{\alpha \beta} (H_1^0-
\overline{H_2^0})+
h^{(2)}_{\alpha \beta} (H_2^0 -\overline{H_1^0}) \right)
z^{\alpha}z^{\beta}
+h.c.\right] \right\} + {\cal O} (z^4) \, ,
\\
V_D & = & \frac{g^2+g'^2}{8}\left( \frac{|H_1^0|^2-|H_2^0|^2}{Y}
\right)^2
+\frac{|H_1^0|^2-|H_2^0|^2}{2Y}(g^2 T_{3L}^{\alpha}-g'^2 Y^{\alpha})
|z^{\alpha}|^2 + {\cal O} (z^4) \, .
\eea
Notice that, for $z^{\alpha}=0$, $V_F$ vanishes identically, and the
same holds for the positive-semidefinite term $V_D$ along the
directions $|H_1^0|=|H_2^0|$. In other words, $\langle V \rangle
\equiv 0$
for $\langle z^{\alpha} \rangle =0$ and for arbitrary values of
$\langle S \rangle$, $\langle T \rangle$, and $\langle |H_1^0|
\rangle =
\langle |H_2^0| \rangle$. Thus the model has a classically
degenerate set of vacua where the cosmological constant vanishes and,
at the same time, both the $SU(2) \times U(1)$ gauge symmetry and
local $N=1$ supersymmetry are spontaneously broken. With the
definitions $s \equiv \langle S+\overline{S} \rangle$, $t \equiv
\langle T+\overline{T} \rangle$ and $x \equiv \langle
H_1^0+\overline{H_2^0}
\rangle$, the gravitino mass reads
\be
m_{3/2}^2 = \langle e^G \rangle = \frac{k^2}{s(t^2-|x|^2)} \, .
\ee

To obtain the physical mass spectrum in the remaining sectors of the
model, one must take into account the presence of non-canonical
kinetic
terms. In particular, the dilaton-axion $S$ and the gauge superfields
can be normalized via a simple rescaling, whereas a non-trivial
mixing
occurs in the $(T,H_1^0,H_2^0)$ sector, due to the more complicated
K\"ahler
metric.

In the gauge boson sector, putting $\rho \equiv \langle |H_1^0|
\rangle
= \langle |H_2^0| \rangle$, we get
\be
m_W^2 = g^2 \frac{\rho^2}{t^2-|x|^2} \, ,
\;\;\;\;\;
m_Z^2 = (g^2+g'^2) \frac{\rho^2}{t^2-|x|^2} \, .
\ee
Observing that, depending on the relative phase of $\langle H_1^0
\rangle$ and
$\langle H_2^0 \rangle$, $0 \le |x|^2 \le 4 \rho^2$, we can see that
it
should
be $\rho^2 / t^2 \simeq 0.25 \times 10^{-32}$ to reproduce the
experimentally measured values of $m_W$ and $m_Z$, irrespectively of
the individual values of $|x|^2$, $\rho^2$ and $t^2$. It is
interesting to observe that, for $\langle H_1^0 \rangle  = - \langle
\ov{H_2^0} \rangle $, one can have $\rho
\ne 0$ with $x  \equiv 0$: this allows for a situation in which the
gravitino mass (and the volume of moduli space)
do not depend on the VEV $\rho$ which breaks $SU(2) \times U(1)$.

In the scalar sector, both spin-0 components of $S$ are massless.
Among the five physical real degrees of freedom of the
$(T,H_1^0,H_2^0)$ sector, which
remain after removing the neutral Goldstone boson associated with the
$Z$, four are massless and one has mass $m_Z$. Observing that the
charged
Higgs fields must appear in the $Y$ function of eq.~(\ref{yyy}) via
the combination $- (H_1^- - \ov{H_2^+} ) (\ov{H_1^-} - H_2^+ )$, it
is easy to verify that the charged Higgs sector contains, besides the
unphysical charged Goldstone boson, a physical state with mass $m_W$.

To discuss the masses of the bosonic and fermionic components of the
`matter' superfields $z^\alpha$, we neglect intergenerational
mixing, rewriting the superpotential couplings involving the neutral
Higgs fields in the simplified form
\be
\label{rewrite}
{1 \over 2} h_{\alpha \beta}^{(1)} z^{\alpha} z^{\beta} H_1^0
+ {1 \over 2} h_{\alpha \beta}^{(2)} z^{\alpha} z^{\beta} H_2^0
= \sum_f h_f f f^c H_1^0 + \sum_{f'} h_{f'} f' f^c{}' H_2^0 \, .
\ee
Then the scalar fields have diagonal masses
\be
m^2_{ff} = m^2_{f^cf^c} = m_{3/2}^2 + m_f^2 \, ,
\;\;\;\;\;
m^2_{f'f'} = m^2_{f^c{}'f^c{}'} = m_{3/2}^2 + m_{f'}^2 \, ,
\ee
where
\be
m_f^2 = \frac{h_f^2 \rho^2}{s(t^2-|x|^2)} \, ,
\;\;\;\;\;
m_{f'}^2 = \frac{h_{f'}^2 \rho^2}{s(t^2-|x|^2)} \, ,
\ee
are the masses of the corresponding fermions. In general, there are
also off-diagonal scalar mass terms of the form
\be
m^2_{ff^c} = \frac{h_f ( \langle H_1^0 \rangle - \langle \ov{H_2^0}
\rangle )}{\sqrt{s (t^2-|x|^2)}} \,
m_{3/2}
\, , \;\;\;\;\;
m^2_{f'f^c{}'} = \frac{h_{f'} ( \langle H_2^0 \rangle - \langle
\ov{H_1^0} \rangle )}{\sqrt{s (t^2-|x|^2)}}
\, m_{3/2}\, .
\ee

The squared mass matrix in the chargino sector reads
\be
\left(
\begin{array}{cc}
m_{3/2}^2 + m_W^2 & m_{3/2} m_W \frac{ \langle H_1^0 \rangle
- \langle \ov{H_2^0} \rangle}{\rho} \\
m_{3/2} m_W \frac{\langle \ov{H_1^0} \rangle - \langle H_2^0
\rangle}{\rho} & m_{3/2}^2 + m_W^2
\end{array}
\right) \, ,
\ee
with eigenvalues $m_{3/2}^2 + m_W^2 \pm m_{3/2}m_W | \langle
H_1^0 \rangle - \langle \ov{H_2^0} \rangle | / \rho$. For
$\langle H_1^0 \rangle =  \langle \ov{H_2^0} \rangle$, one
gets two degenerate states of mass $\sqrt{m_{3/2}^2 + m_W^2}$,
whereas the maximum mass splitting is obtained for $\langle
H_1^0 \rangle = - \langle \ov{H_2^0} \rangle$, in which case
the two masses are $|m_{3/2} \pm m_W|$.

In the neutralino sector (which includes here one more physical
state than in the MSSM), a linear combination of the fermionic
$S$ and $(T,H_1^0,H_2^0)$ fields can be identified with the
goldstino. The three remaining physical states mix between them
and with the two neutral electroweak gauginos (the gluino mass is
equal to the gravitino mass). The corresponding masses satisfy
the sum rule $\sum_{i=1}^5 m_i^2 = 5 m_{3/2}^2 + 2 m_Z^2$.

To understand the structure of the present model better, it is
convenient to take the limit $\rho / t \rightarrow 0$, which
leads to a conventional supergravity model with hidden sector
and, when interactions of gravitational strength are neglected,
to a special version of the MSSM. In such a limit, the goldstino
becomes a linear combination of the $S$ and $T$ fermions only,
with mixing angle, in the notation of [\ref{bim}], $\sin^2\theta
= 1/3$. The orthogonal combination has mass $m_{3/2}$. The MSSM
mass parameters take the special values
\be
\label{masspar}
\begin{array}{ccc}
m_{1/2}^2 = m_{3/2}^2 \, ,
&
m_0^2({\rm matter}) = m_{3/2}^2 \, ,
&
m_0^2({\rm Higgs}) = - m_{3/2}^2 \, ,
\\ & & \\
\mu^2 = m_{3/2}^2 \, ,
&
A^2 = m_{3/2}^2 \, ,
&
B=0 \, ,
\end{array}
\ee
as can be easily checked by looking at the limiting form of the
supergravity mass matrices, remembering that in the chosen limit
the canonically normalized Higgs fields are given by $H_{1,2} / t$,
and the MSSM Yukawa couplings by $h_{f,f'} / \sqrt{s}$. Notice in
particular that the MSSM mass terms exhibit remarkable universality
properties, much more stringent than usually assumed in the general
MSSM framework, with one important exception: since the kinetic terms
for the Higgs and matter fields have different scaling properties
with
respect to the $t$ modulus, the corresponding soft scalar masses have
different values. In particular, the standard mass parameters of the
classical MSSM Higgs potential are given by $m_1^2 = m_2^2 = m_3^2 =
0$, which allows for $SU(2) \times U(1)$ breaking along the flat
direction $|H_1^0 | = |H_2^0|$.

\vspace{1cm}
{\bf 3.}
In conclusion, we have constructed a semi-realistic supergravity
model in which, already at the classical level, both $N=1$
supersymmetry and the $SU(2) \times U(1)$ gauge symmetry are
spontaneously broken with naturally vanishing vacuum energy,
thanks to the remarkable geometrical properties of superstring
effective supergravities.

We would like to stress again some important differences with the
supergravity models considered so far in the literature. Usually,
$SU(2) \times U(1)$ is unbroken at the classical level, and the
goldstino is neutral under $SU(2) \times U(1)$: to break $SU(2)
\times U(1)$ one appeals to radiative corrections. In the present
model, $SU(2) \times U(1)$ breaking occurs already at the
classical level, and the goldstino has non-vanishing components
along the neutral higgsino directions.

Due to the role played by the internal singlet modulus $T$, which
takes part in the superhiggs mechanism and forces the ratio $\rho
/t$ to be of order $m_Z / \mpl$, in order to reproduce the
experimental value of the electroweak scale, the interactions of
the gravitino via its goldstino (higgsino) components are suppressed
down to gravitational strength. However, it is conceivable that one
could construct similar models where $S$ and $T$ do not take part
in the superhiggs mechanism, and all the light mass eigenstates
have interactions of electroweak strength. In such
a framework, some cosmological problems of gravitationally-coupled
states with electroweak-scale masses would be avoided (even if others
could arise). We have not yet been able to build a model with these
properties. To do so, one should probably move to models where the
spontaneously broken gauge group $SU(2) \times U(1)$ can be embedded
into the `compensator' subgroup of the isometry group of the manifold
containing the Higgs fields. The manifold $SO(2,n)/[SO(2) \times
SO(n)]$ is not a viable candidate, whereas manifolds such as
$SU(3,n)/[SU(3) \times SU(n) \times U(1)]$ or $SU(2,n)/ [SU(2)
\times SU(n) \times U(1)]$, associated to some untwisted sectors
of $Z_3$ and $Z_6$ orbifold models, might be viable\footnote{We
thank C.~Kounnas for a clarifying discussion on this point.}.

The tree-level potential of our model exhibits several flat
directions, associated to classically massless scalar fields,
some of which control, via their VEVs, the gauge- and
supersymmetry-breaking scales and the dimensionless couplings.
Generically, we expect these flat directions to be removed by
the consistent inclusion of quantum corrections, which
should fix the VEVs along the flat directions and the corresponding
scalar masses. However, it is difficult to make definite statements
about the actual values of these VEVs and masses, since expectations
based on simple dimensional arguments may turn out to be incorrect.
For example, in the models of ref.~[\ref{lhc}] the singlet scalar
partners of the
goldstino have masses of order $m_{3/2}^2/\mpl$, and not $m_{3/2}$,
because of the absence of loop contributions to the vacuum energy of
order $m_{3/2}^2 \mpl^2$. Similarly, if loop corrections fix $m_W$ to
its experimental value, and CP is not spontaneously broken, loop
contributions to the lightest CP-even Higgs boson mass associated
with
top and stop loops [\ref{erz}] are of order $m_t^2/m_W$, which
already
brings the latter into the phenomenologically acceptable region. It
might
even be that some direction remains flat, as assumed for example in
the
recent work by Polyakov and Damour [\ref{pta}], who discussed
possible phenomenological implications of a massless dilaton.
Consistent inclusion of perturbative quantum corrections would
require the knowledge not only of the singlet-moduli dependence,
but also of the Higgs-fields dependence of the string loop
corrections
to the low-energy effective supergravity. This is not fully available
yet, but it is not inconceivable that it will be soon, at least in
some
simple classes of four-dimensional string constructions. In this
case,
many more interesting questions could be addressed, including a
quantitative study of the dynamical determination of the gravitino
and electroweak scales.

\vfill{
\section*{Acknowledgements}
A.B. would like to thank the CERN Theory Division for the kind
hospitality during part of this work. His work was partially
supported by the U.S. Department of Energy under Contract
DE-AC03-76SF00098. F.Z. would like to thank: the ITP at Santa
Barbara for the kind hospitality during the initial stage of
this work; L.~Randall for arousing his spirit of contradiction;
S.~Ferrara and C.~Kounnas for many instructive and entertaining
discussions.}

\newpage
\section*{References}
\begin{enumerate}
\item
\label{susy}
For reviews and references, see, e.g.:
\\
H.-P. Nilles, Phys. Rep. 110 (1984) 1;
\\
S. Ferrara, ed., `Supersymmetry' (North-Holland, Amsterdam, 1987);
\\
F.~Zwirner, in `Proceedings of the 1991 Summer School in High Energy
Physics and Cosmology', Trieste, 17 June--9 August 1991 (E.~Gava,
K.~Narain, S.~Randjbar-Daemi, E.~Sezgin and Q.~Shafi, eds.), Vol.~1,
p.~193.
\item
\label{weinberg}
S.~Weinberg, Rev. Mod. Phys. 61 (1989) 1.
\item
\label{nscl}
E.~Cremmer, S.~Ferrara, C.~Kounnas and D.V.~Nanopoulos,
Phys. Lett. B133 (1983) 61;
\\
R.~Barbieri, S.~Ferrara and E.~Cremmer, Phys. Lett. B163 (1985) 143.
\item
\label{nsqu}
J.~Ellis, A.B.~Lahanas, D.V.~Nanopoulos and K.~Tamvakis, Phys. Lett.
B134 (1984) 429;
\\
J.~Ellis, C.~Kounnas and D.V.~Nanopoulos,
Nucl. Phys. B241 (1984) 406 and B247 (1984) 373.
\item
\label{lhc}
S.~Ferrara, C.~Kounnas and F.~Zwirner, preprint CERN-TH.7192/94,
LPTENS-94/12, UCLA/94/TEP13, hep-th/9405188, to appear in Nucl.
Phys.~B.
\\
See also S.~Ferrara, C.~Kounnas, M.~Porrati and F.~Zwirner,
Phys. Lett. B194 (1987) 366.
\item
\label{kpz}
C.~Kounnas, I.~Pavel and F.~Zwirner, Phys. Lett. B335 (1994) 403.
\item
\label{ss}
C.~Kounnas and M.~Porrati, Nucl. Phys. B310 (1988) 355;
\\
S.~Ferrara, C.~Kounnas, M.~Porrati and F.~Zwirner,
Nucl. Phys. B318 (1989) 75;
\\
M.~Porrati and F.~Zwirner, Nucl. Phys. B326 (1989) 162;
\\
C.~Kounnas and B.~Rostand, Nucl. Phys. B341 (1990) 641;
\\
I.~Antoniadis, Phys. Lett. B246 (1990) 377;
\\
I.~Antoniadis and C.~Kounnas, Phys. Lett. B261 (1991) 369;
\\
I.~Antoniadis, C.~Mu\~noz and M.~Quir\'os, Nucl. Phys. B397 (1993)
515.
\item
\label{strqu}
L.E.~Ib\`a\~nez and H.P.~Nilles, Phys. Lett. B169 (1986) 354;
\\
H.~Itoyama and J.~Leon, Phys. Rev. Lett. 56 (1986) 2352;
\\
E.~Martinec, Phys. Lett. B171 (1986) 2352;
\\
M.~Dine and N.~Seiberg, Phys. Rev. Lett. 57 (1986) 2625;
\\
H.P.~Nilles, Phys. Lett. B180 (1986) 240;
\\
M.~Dine, N.~Seiberg and E.~Witten, Nucl. Phys. B289 (1987) 589;
\\
V.S.~Kaplunovsky, Nucl. Phys. B307 (1988) 145;
\\
L.J.~Dixon, V.S.~Kaplunovsky and J.~Louis, Nucl. Phys. B355 (1991)
649;
\\
J.-P.~Derendinger, S.~Ferrara, C.~Kounnas and F.~Zwirner, Nucl. Phys.
B372 (1992) 145 and Phys. Lett. B271 (1991) 307;
\\
G.~Lopes Cardoso and B.A.~Ovrut, Nucl. Phys. B369 (1992) 351;
\\
I.~Antoniadis, K.S.~Narain and T.~Taylor, Phys. Lett. B276 (1991) 37
and Nucl. Phys. B383 (1992) 93;
\\
I.~Antoniadis, E.~Gava, K.S.~Narain and T.R.~Taylor, Nucl. Phys. B407
(1993) 706.
\item
\label{gcond}
J.-P.~Derendinger, L.E.~Ib\'a\~nez and H.P.~Nilles, Phys. Lett. B155
(1985) 65;
\\
M.~Dine, R.~Rohm, N.~Seiberg and E.~Witten, Phys. Lett. B156 (1985)
55;
\\
C.~Kounnas and M.~Porrati, Phys. Lett. B191 (1987) 91;
\\
A.~Font, L.E.~Ib\'a\~nez, D.~L\"ust and F.~Quevedo, Phys. Lett.
B245 (1990) 401;
\\
S.~Ferrara, N.~Magnoli, T.R.~Taylor and G.~Veneziano, Phys.
Lett. B245 (1990) 409;
\\
H.-P.~Nilles and M.~Olechowski, Phys. Lett. B248 (1990) 268;
\\
P.~Bin\'etruy and M.K.~Gaillard, Phys. Lett. B253 (1991) 119;
\\
V.~Kaplunovsky and J.~Louis, Nucl. Phys. B422 (1994) 57;
\\
J.H.~Horne and G.~Moore, Yale preprint YCTP-P2-94, hep-th/9403058.
\item
\label{others}
D.~Amati, K.~Konishi, Y.~Meurice, G.C.~Rossi and G.~Veneziano, Phys.
Rep. 162 (1988) 169, and references therein;
\\
A.E.~Nelson and N.~Seiberg, Nucl. Phys. B416 (1994) 46, and
references therein.
\item
\label{cosmo}
H.~Pagels and T.R.~Primack, Phys. Rev. Lett. 48 (1982) 223;
\\
S.~Weinberg, Phys. Rev. Lett. 48 (1982) 1303;
\\
J.~Preskill, M.B.~Wise and F.~Wilczek, Phys. Lett. B120 (1983) 127;
\\
L.P.~Abbott and P.~Sikivie, Phys. Lett. B120 (1983) 133;
\\
M.~Dine and W.~Fischler, Phys. Lett. B120 (1983) 137;
\\
L.~Krauss, Nucl. Phys. B227 (1983) 556;
\\
G.D.~Coughlan, W.~Fischler, E.W.~Kolb, S.~Raby and G.G.~Ross,
Phys. Lett. B131 (1983) 59;
\\
J.~Ellis, D.V.~Nanopoulos and M.~Quir\'os, Phys. Lett. B174 (1986)
176;
\\
J.~Ellis, N.C.~Tsamis and M.~Voloshin, Phys. Lett. B194 (1987) 291;
\\
K.~Rajagopal, M.~Turner and F.~Wilczek, Nucl. Phys. B358 (1991) 447;
\\
R.~Brustein and P.J.~Steinhardt, Phys. Lett. B302 (1993) 196;
\\
B.~de~Carlos, J.A.~Casas, F.~Quevedo and E.~Roulet, Phys. Lett. B318
(1993) 447;
\\
T.~Banks, D.B.~Kaplan and A.E.~Nelson, Phys. Rev. D49 (1994) 779.
\\
W.~Fischler, Phys. Lett. B322 (1994) 277;
\\
L.~Randall and S.~Thomas, preprint MIT-CTP-2331, NSF-ITP-94-70, SCIPP
94-16,
hep-ph/9407248.
\item
\label{quiros}
C.~Kounnas, J.~Leon and M.~Quir\'os, Phys. Lett. B129 (1983) 67;
\\
C.~Kounnas, D.V.~Nanopoulos and M.~Quir\'os, Phys. Lett. B129
(1983) 223.
\item
\label{fayet}
P.~Fayet, Phys. Lett. 70B (1977) 461.
\item
\label{efft}
E.~Witten, Phys. Lett. B155 (1985) 151;
\\
S.~Ferrara, C.~Kounnas and M.~Porrati, Phys. Lett. B181 (1986) 263;
\\
S.~Ferrara, L.~Girardello, C.~Kounnas and M.~Porrati, Phys. Lett.
B192 (1987) 368 and B194 (1987) 358;
\\
I.~Antoniadis, J.~Ellis, E.~Floratos, D.V.~Nanopoulos and T.~Tomaras,
Phys. Lett. B191 (1987) 96;
\\
M.~Cvetic, J.~Louis and B.~Ovrut, Phys. Lett. B206 (1988) 227;
\\
S.~Ferrara and M.~Porrati, Phys. Lett. B216 (1989) 1140;
\\
M.~Cvetic, J.~Molera and B.~Ovrut, Phys. Rev. D40 (1989) 1140;
\\
L.~Dixon, V.~Kaplunovsky and J.~Louis, Nucl. Phys. B329 (1990) 27.
\item
\label{fds}
For a review and references see, e.g.:
\\
B.~Schellekens, ed., `Superstring construction' (North-Holland,
Amsterdam, 1989).
\item
\label{fkpz}
S.~Ferrara, C.~Kounnas, M.~Porrati and F.~Zwirner, as in
ref.~[\ref{ss}].
\item
\label{ferm}
H.~Kawai, D.C.~Lewellen and S.-H.H. Tye, Nucl. Phys. B288 (1987) 1;
\\
I.~Antoniadis, C.~Bachas and C.~Kounnas, Nucl. Phys. B289 (1987) 87.
\item
\label{orb}
L.~Dixon, J.~Harvey, C.~Vafa and E.~Witten, Nucl. Phys. B261 (1985)
678;
\\
K.S.~Narain, M.H.~Sarmadi and C.~Vafa, Nucl. Phys. B288 (1987) 551.
\item
\label{param}
I.~Antoniadis, E.~Gava, K.S.~Narain and T.R.~Taylor, preprint
NUB-3084,
IC/94/72, CPTH-A282.0194, hep-th/9405024;
\\
G.~Lopes~Cardoso, D.~L\"ust and T.~Mohaupt, preprint HUB-IEP-94/6,
hep-th/9405002.
\item
\label{bim}
A.~Brignole, L.E.~Ib\'a\~nez and C.~Mu\~noz, Nucl.
Phys. B422 (1994) 125.
\item
\label{erz}
J. Ellis, G. Ridolfi and F. Zwirner, Phys. Lett. B257 (1991) 83
and Phys. Lett. B262 (1991) 477;
\\
Y. Okada, M. Yamaguchi and T. Yanagida, Prog. Theor. Phys. Lett.
85 (1991) 1 and Phys. Lett B262 (1991) 54;
\\
H.E. Haber and R. Hempfling, Phys. Rev. Lett. 66 (1991) 1815;
\\
R. Barbieri, M. Frigeni and M. Caravaglios, Phys. Lett. B258 (1991)
167.
\item
\label{pta}
T.~Damour and A.M.~Polyakov, Nucl. Phys. B423 (1994) 532.
\end{enumerate}
\end{document}